\documentclass[prd,preprint,floatfix,eqsecnum,nofootinbib,12pt]{revtex4}
\usepackage{hyperref}
\usepackage{amssymb,amsmath,amsthm,graphicx,ulem,slashed}
\pdfoutput=1
\usepackage{graphicx,subfigure}
\usepackage{cleveref}
\usepackage{ulem}
\usepackage{epsfig}
\usepackage{amsmath}
\usepackage{amsfonts}
\usepackage{amssymb}
\usepackage[usenames]{color}
\usepackage[letterpaper,left=2cm,right=2cm,top=2.32cm,bottom=2.32cm]{geometry}
\usepackage[T1]{fontenc}

\newcommand{\cD}{\ensuremath{\mathcal{D}}}

\def\be{\begin{equation}}
\def\ee{\end{equation}}

\newcommand {\cL}{{\cal L}}

\newcommand {\cN}{{\cal N}}
\newcommand {\cO}{{\cal O}}

\def\a{\alpha}
\def\b{\beta}

\def\d{\delta}
\def\e{\epsilon}
\def\f{\phi}
\def\g{\gamma}

\def\l{\lambda}

\def\q{\theta}

\def\s{\sigma}

\def\z{\zeta}

\def\F{\Phi}

\def\U{\Upsilon}

\def\ri{{\rm i}}
\def\re{{\rm e}}

\newcommand{\ad}{{\dot{\alpha}}}                 
\newcommand{\bd}{{\dot{\beta}}}

\newcommand{\pa}{\partial}                

\newcommand{\bea}{\begin{eqnarray}}
\newcommand{\eea}{\end{eqnarray}}
\newcommand{\non}{\nonumber}
\newcommand{\ba}{\begin{array}}
\newcommand{\ea}{\end{array}}

\def\double #1{#1{\hbox{\kern-2pt $#1$}}}

\newcommand{\bsubeq}{\begin{subequations}}
\newcommand{\esubeq}{\end{subequations}}

\newcommand{\rd}{\mathrm d}
\begin{document}

\title{{\LARGE {Supersymmetric dS/CFT}}}

\author{Thomas Hertog}
\email{thomas.hertog@kuleuven.be}
\affiliation{ Institute for Theoretical Physics, KU Leuven, Celestijnenlaan 200D, 3001 Leuven, Belgium}
\author{Gabriele Tartaglino-Mazzucchelli}
\email{gabriele.tartaglino-mazzucchelli@kuleuven.be}
\affiliation{ Institute for Theoretical Physics, KU Leuven, Celestijnenlaan 200D, 3001 Leuven, Belgium}
\author{Thomas Van Riet}
\email{thomas.vanriet@kuleuven.be}
\affiliation{ Institute for Theoretical Physics, KU Leuven, Celestijnenlaan 200D, 3001 Leuven, Belgium}
\author{Gerben Venken}
\email{gerben.venken@kuleuven.be}
\affiliation{ Institute for Theoretical Physics, KU Leuven, Celestijnenlaan 200D, 3001 Leuven, Belgium}

\begin{abstract}

We put forward new explicit realisations of dS/CFT that relate ${\cal N}=2$ supersymmetric Euclidean vector models  with reversed spin-statistics in three dimensions to specific supersymmetric Vasiliev theories in four-dimensional 
de Sitter space. The partition function of the free supersymmetric vector model deformed by a range of low spin deformations that preserve supersymmetry appears to specify a well-defined wave function with asymptotic de Sitter boundary conditions in the 
bulk. In particular we find the wave function is globally peaked at undeformed de Sitter space, with a low amplitude for strong 
deformations. This suggests that supersymmetric de Sitter space is stable in higher-spin gravity and in particular free from ghosts. 
We speculate this is a limiting case of the de Sitter realizations in exotic string theories.

\end{abstract}

\maketitle

\tableofcontents

\newpage

\section{Introduction}

Gauge-gravity duality with de Sitter (dS) boundary conditions
 \cite{Hull:1998vg,Balasubramanian2001,Strominger:2001pn}
  has proved to be a fruitful route to put cosmology on firm theoretical ground. 
  In its most ambitious and fundamental form, dS/CFT conjectures that the partition function of certain deformations of 
  three dimensional Euclidean CFTs yields a precise formulation of the Hartle-Hawking wave function of the universe 
  \cite{Hartle:1983ai}. Since a wave function of the universe specifies a prior for cosmology, which in turn determines 
  the theory's predictions for cosmological observations, this potentially provides a solid foundation for cosmology. 
Schematically and in the large three-volume regime the proposed dual form of the wave function reads 
\be
\Psi_{HH} [h_{ij}, A_s]= Z_{QFT}[\tilde h_{ij}, J_s] \exp(\ri S_{st}[h_{ij}, A_s]/\hbar)   \ .
\label{dSCFT}
\ee
Here $A_s$ stands for the matter configurations of spin $s$ and $h_{ij}$ is the three-geometry of the spacelike 
surface $\Sigma$ on which $\Psi$ is evaluated. In this paper we take the latter to be topologically a three-sphere. 
The sources $(\tilde h_{ij}, J_s)$ in \eqref{dSCFT} are conformally related to the argument $(h_{ij}, A_s)$ of the wave 
function, and $S_{st}$ are the usual surface terms.

It is a central question in holographic cosmology what class of deformed CFTs in \eqref{dSCFT} specifies a 
well-defined, normalizable wave function. Euclidean AdS/CFT provides a starting point to study this since its generalization to complex relevant deformations of CFTs implies a realisation of dS/CFT that is valid in the semiclassical approximation 
in Einstein gravity \cite{Maldacena:2002vr,Harlow2011,Maldacena2011,Hertog2011} and possibly exact in Vasiliev 
gravity in dS \cite{Anninos:2011ui}. 
It has been suggested indeed that Euclidean AdS and Lorentzian dS, and their duals, can be viewed as two real 
domains of a single complexified theory 
\cite{Maldacena:2002vr,Hull:1998vg,Dijkgraaf:2016lym,Skenderis:2007sm,Bergshoeff:2007cg,Hartle2012b}. 

In these examples of dS/CFT the partition functions featuring in \eqref{dSCFT} are the inverse of those of the original 
AdS/CFT duals \cite{Hertog2011}. The case of higher-spin gravity is particularly illuminating because the duals are 
vector models for which the partition function can be evaluated explicitly for a range 
of deformations\footnote{This amounts to a minisuperspace approximation in cosmology.
Note that we will be using the term superspace in two different contexts: on the one hand there is the minisuperspace
of quantum cosmology, on the other hand we have the superspace used for supersymmetry.} 
\cite{Anninos:2012ft,Anninos:2013rza,Bobev:2016sap,Conti:2017pqc}. The Vasiliev higher-spin (HS) theory has massive 
scalars and an infinite tower of massless gauge fields of increasing spin \cite{Vasiliev:1990en}. The duals have 
conserved currents for the same symmetries \cite{Klebanov:2002ja,Giombi:2016ejx}. Deforming the boundary theory 
action with a conserved current $J_s$ corresponds to turning on the spin-$s$ field $A_s$
 in the bulk.\footnote{The spin-$0$ field carries no indices and is not really a current. Still, it is usually referred 
 to as such.} Explicit calculations of the partition function as a function of homogeneous scalar and spin-2 
 deformations in the $Sp(N)$ vector model, which is dual to the minimal Vasiliev theory in dS \cite{Anninos:2011ui}, 
 have provided some evidence that dS/CFT yields a well-defined wave function and in particular one which is better
  behaved than the usual semiclassical Hartle-Hawking wave function in Einstein gravity \cite{Hawking:2017wrd}.

So far attention has focussed on the duality between the minimal Vasiliev theory in de Sitter and the dual $Sp(N)$ 
models consisting of anti-commuting scalars. In this case the dS duality follows relatively directly from the higher-spin 
duality in AdS \cite{Giombi:2016ejx}. However, the connection between the AdS and the dS domain of the theories 
appears more general and profound \cite{Hull:1998vg,Dijkgraaf:2016lym,Bergshoeff:2007cg,Hartle2012b}, which 
suggests there may well be a broader set of realizations of dS/CFT in the same spirit. Here we show that the 
supersymmetric extension of the higher-spin duality in AdS \cite{Chang:2012kt} also carries over to dS. 
This is particularly interesting because a supersymmetric realisation of dS/CFT has a better chance of specifying a 
well-defined wave function that predicts stable asymptotic de Sitter space. In fact Ooguri and Vafa
 \cite{Ooguri:2016pdq} have recently advocated that supersymmetry is essential even for 
AdS holography,\footnote{Related to this, Danielsson et al. \cite{Danielsson:2016mtx} have conjectured that all 
non-supersymmetric AdS vacua are even perturbatively unstable, and in \cite{Freivogel:2016qwc} it was argued that 
all non-supersymmetric vacua -- be it AdS, Minkowski or dS -- must eventually decay.} although they note higher-spin 
gravity may evade their arguments.

It is usually argued that unbroken supersymmetry and dS space do not go together 
(see e.\,g.~\cite{Witten:2001kn}).\footnote{Note that superconformal theories on 
a fixed de Sitter background do exist \cite{Anous:2014lia}. They avoid the argument of 
\cite{Witten:2001kn} by the existence of an everywhere timelike conformal killing vector.}
This is because in dS space there is no positive conserved quantity, whereas supersymmetry would allow one to 
construct one. Indeed if there were a nonzero supercharge $Q$ then either 
$(Q + Q^\dagger)$ or $\ri(Q - Q^\dagger)$ 
would be Hermitian. Redefining $Q$ to be the Hermitian supercharge would then imply that $Q^2$ were a positive 
conserved quantity, which does not exist in dS.

The fact that the known dS vacua in supergravity have ghosts \cite{Pilch:1984aw,Lukierski:1984it}, 
indicating they are 
perturbatively unstable, is a manifestation of this general argument. Supersymmetric HS gravity theories 
in de Sitter may however circumvent this problem. This is because in the higher-spin theories in dS constructed in 
\cite{Sezgin:2012ag} the Hermitian conjugate is an anti-involution, defined as $(Q^\dagger)^\dagger = - Q$. With this 
definition one cannot construct a Hermitian quantity from a supercharge $Q$. In fact, the $\cN=2$ ${\rm dS}_4$ 
supersymmetry algebra is realized in terms of oscillators of Vasiliev theory, comprising the bosonic fields $y_\a$ and 
$\bar{y}_\ad=(y_\a)^\dagger$, such that the momentum operator $P_a=\ri/4(\s_a)^{\a\bd}y_\a\bar{y}_\bd$ is 
anti-Hermitian. Hence, from the irrepresentations of the higher-spin superalgebra there does not seem to exist an 
operator associated to a globally defined Hamiltonian in the first place. 

In this paper we provide evidence, using holography, that supersymmetric de Sitter space is stable and has no ghosts 
in higher-spin gravity. We first propose a supersymmetric generalization of the higher-spin dualities in de Sitter. 
The bulk theories involved are the supersymmetric extensions of Vasiliev theory described in \cite{Sezgin:2012ag}. 
On the boundary side we construct, in Section \ref{sec:deform}, new ${\cal N}=2$ supersymmetric extensions of the 
three-dimensional $Sp(N)$ models. We then relate these to the theories of Sezgin and Sundel in Section \ref{dual}, thereby 
establishing a supersymmetric gauge-gravity duality with de Sitter boundary conditions. We also briefly discuss the fermions in the theory. In Section \ref{partition-function} 
we evaluate the partition function of the supersymmetric extension of the free $Sp(N)$ model as a function of 
homogeneous scalar, vector and spin-2 deformations that preserve supersymmetry. The duality put forward in Section \ref{sec:deform} conjectures that the partition function specifies the Hartle-Hawking wave function in a supersymmetric minisuperspace consisting of anisotropic deformations of de Sitter space with scalar and vector matter. We find the wave function is globally peaked at the undeformed de Sitter space, with a low amplitude for strong deformations. This indicates that supersymmetric de Sitter space is stable in higher-spin gravity. 

It is tempting to speculate that our findings are connected to the supersymmetric dS constructions in exotic string 
theories \cite{Hull:1998vg}. The latter have vector ghosts in their supergravity limits related to the existence of  
non-compact $R$-symmetry groups in their representation of the 
algebra.\footnote{This differs from the $\cN=2$ 
Vasiliev case in \cite{Sezgin:2012ag} where the dS $R$-symmetry group is 
$SO(2)_R\simeq U(1)_R$. The $\cN=2$ case is actually the only dS supergroup
possessing compact $R$-symmetry \cite{Pilch:1984aw,Lukierski:1984it}.} 
However Hull has argued that the massive string states in exotic string theories may well 
render the de Sitter vacua ghost-free and unitary. In Section \ref{sec:string} we conjecture that the supersymmetric 
higher-spin theories in dS that we construct are related indeed to the tensionless limit of these exotic string theories. 

\section{Supersymmetric Vector Models and Duality}
\label{sec:deform}

\subsection{Free chiral superfields and higher-spin deformations}

To construct Euclidian supersymmetric vector models we must consider complex fields and $U(N)$ global symmetry
instead of $O(N)$ or $Sp(N)$. The free $U(N)$ vector model is a theory of $N$ commuting complex scalar fields 
$\f^i$ and $\tilde{\f}_i$, with $i=1,\ldots,N$, transforming in the fundamental and anti-fundamental representations of 
$U(N)$ with Lagrangian $\cL= \pa^\mu \tilde{\f}_i\pa_\mu\f^i$. The $Sp(N)$ model or more precisely the
anti-commuting $U(N)$ model, also referred to as the $U(-N)$ model, is defined by changing the statistics of the 
fields. This yields a set of complex anti-commuting scalar fields $\varphi^i$ and $\tilde{\varphi}_i$ governed by the 
Lagrangian $\cL= \pa^\mu \tilde{\varphi}_i\pa_\mu\varphi^i$.

Here we are interested in the three-dimensional, Euclidean $\cN=2$ supersymmetric extension of these vector 
models.\footnote{See e.g.\,\cite{Leigh:2003gk,Chang:2012kt,Hikida:2017ecj,Honda:2017nku} for an analysis of the 
analogous extension in the context of the AdS-Vasiliev/CFT duality.} Three dimensional $\cN=2$ supersymmetry has 
four real supercharges. 
The supersymmetric extension of a 3D scalar field can be described off-shell by using chiral and anti-chiral 
superfields $\F^i$ and $\tilde{\F}_i$, which we assume to be commuting for now. These satisfy the constraints
 $\tilde{D}_\a\F^i=0$ and $D_\a\tilde{\F}_i=0$ where the $\cN=2$ superspace coordinates and spinor covariant 
 derivatives are denoted $z^M=(x^\mu,\theta^\a,\tilde{\theta}^\a)$
  and $(D_\a,\tilde{D}_\a)$.\footnote{We adopt  the notation of  \cite{Closset:2012ru,Willett:2016adv} except for their
 supersymmetry parameters $(\z,\tilde{\z})$ which we will name $(\e,\tilde{\e})$.}
Note that in Euclidian signature  $\tilde{\F}_i$ is not necessarily the complex conjugate of $\F^i$.
The off-shell components of the chiral superfield $\F^i$ are $(\f^i,\psi_\a^i,F^i)$ with 
$\f^i(x)= \F^i|_{\q=0}$ the physical complex scalar fields, 
$\psi_\a^i(x)=1/\sqrt{2}D_\a\F^i|_{\q=0}$ Dirac spinors and 
$F^i(x)=-1/4D^2\F^i|_{\q=0}$ complex auxiliary fields (and analogously for the anti-chiral multiplet).
Below we give the supersymmetry transformations for the fields in a chiral multiplet.
The $\cN=2$ supersymmetric $U(N)$ model is then 
\bea
S=
\int\rd^3x\rd^4\q
\,\tilde\F_i\F^i
=\int\rd^3x\Big{[}
\pa_\mu \tilde{\phi}_i \pa^\mu \phi^i
- \ri \tilde{\psi}_i \slashed{\pa} \psi^i
- \tilde{F}_i F^i
\Big{]}
~,
\label{free-U(N)}
\eea
where the first term is the free bosonic $U(N)$ model, the second term describes the free massless 
Gross-Neveu model
and the last term is non-dynamical. If the $U(1)_R$ charge and the conformal dimension of $\F^i$ are chosen 
to be 1/2, this action is invariant under the 3D Euclidian superconformal group $OSp(2|2,2)$.\footnote{We use
the $OSp(2|2,2)$ supergroup notation employed in \cite{Kuzenko:2014yia} 
which differs from the unitary quaternionic one used in \cite{Pilch:1984aw,Lukierski:1984it}.
In the latter case the minimal 3D Euclidian superconformal group, equivalent to the minimal 4D dS supergroup,
is denoted as $UU_\a(1,1;1;\mathbb{H})$.}
Moreover, the action possesses an infinite set of singlet higher-spin supercurrents 
$J^{(s)}$ \cite{Nizami:2013tpa}. The scalar supercurrent of the free theory is given by $J^{(0)}:=\tilde\F_i \F^i$ 
and satisfies on-shell the conservation equation $D^2 J^{(0)}=\tilde{D}^2  J^{(0)}=0$.

The superspace approach straightforwardly allows us to define an $\cN=2$ extension of the $U(-N)$ model: It suffices 
to replace the commuting chiral and anti-chiral superfields $\F^i$ and $\tilde{\F}_i$, with a set of anti-commuting fields 
$\U^i$ and $\tilde{\U}_i$. The change of statistics is such that\footnote{Note also that the commutation rules with the 
spinor derivatives are somewhat subtle, e.g. $D_\a{\U}_i\U_j=(D_\a{\U}_i)\U_j-{\U}_iD_\a\U_j$.} $\U^i\U^j=-\U^j\U^i$, 
$\tilde\U_i\tilde\U_j=-\tilde\U_j\tilde\U_i$, $\U^i\tilde\U_j=-\tilde\U_j\U^i$. The anti-commuting chiral superfields 
have components given by the physical anti-commuting complex scalars $\varphi^i(x)$, 
commuting fermions $\chi_\a^i(x)$ and anti-commuting auxiliary scalar fields $G^i(x)$.
The supersymmetry transformations of $\U^i$ and $\tilde{\U}_i$ are the same as those of $\F^i$ and $\tilde{\F}_i$. 
The only difference is that the component fields in the off-shell multiplet now have reversed statistics.
The full superspace Lagrangian for the $U(-N)$ model is then simply given by $\tilde\U_i\U^i$, as in \eqref{free-U(N)}.
For dimension half anti-commuting chiral multiplets this has the same $OSp(2|2,2)$ symmetry and higher-spin 
currents as the $\cN=2$ $U(N)$ theory.

One can extend the free $U(-N)$ model to include interactions by considering an arbitrary full superspace 
Lagrangian\footnote{This is the analogue of the K\"ahler potential for a standard supersymmetric sigma-model.} 
$K(\U^i,\tilde{\U}_i)$, and chiral and anti-chiral superpotentials $W(\U^i)$ and $\tilde{W}(\tilde{\U}_i)$.
For instance an extension we will consider is the free $\cN=4$ hypermultiplet $U(-N)$ model. This is defined by 
a full superspace Lagrangian $(\tilde\U_+{}_i\U_+^i + \tilde\U_-^i\U_-{}_i)$ involving two sets of free anti-commuting 
chiral superfields  $\U_{+}^i$ and $\U_{-}{}_i$ and the antichiral cousins. 
By using the chiral composite $C=\U_+^i\U_-{}_i$ 
one can construct $U(N)$ invariant superpotentials $W(C)$. The quartic superpotential $(\U_+^i\U_-{}_i)^2$ provides 
an example and represents a classically marginal deformation of the $U(-N)$ model preserving the full $\cN=2$ 
superconformal group, but not the higher-spin (HS) symmetry.

We do not elaborate further on interacting theories but concentrate first on the free $\cN=2$ $U(-N)$ model defined on 
the three sphere and deformed by background boundary sources for higher-spin supercurrents. This can be used to 
realize the simplest example of a supersymmetric extension of dS/CFT. In this context, the boundary sources are 
related to the argument of the bulk wave function in the large volume limit (cf. \eqref{dSCFT}).

Consider background $\cN=2$ conformal higher-spin superfields \cite{Kuzenko:2016bnv}
$H^{(s)} :=H_{\alpha_1\alpha_2\cdots\alpha_{2s}}(x,\q,\tilde{\q})$, 
together with the  primary supercurrents of the free chiral multiplets \cite{Nizami:2013tpa},
$J^{(s)} :=J_{\alpha_1\alpha_2\cdots\alpha_{2s}}(x,\q,\tilde{\q})$.
These are  both completely symmetric tensors in the $2s$ spinor indices.
Then consider a linearly deformed full superspace Lagrangian of the form
 \bea
\cL[H^{(s)}]=\tilde\U_i\U^i + \sum_{s=0}^{+\infty} H^{\alpha_1\alpha_2\cdots\alpha_{2s}}
J_{\alpha_1\alpha_2\cdots\alpha_{2s}}
~.
\label{Lagrangian-HS}
\eea
The deformed model is also off-shell $OSp(2|2,2)$ invariant as can be proven by using the results of 
\cite{Kuzenko:2016bnv} (see also the recent 4D analysis of \cite{Kuzenko:2017ujh}).
The partition function of the Lagrangian \eqref{Lagrangian-HS}
leads to a natural superfield description of the Hartle-Hawking wave function \eqref{dSCFT}.
The spin zero supercurrent is $J^{(0)}:=\tilde{\U}_i\U^i$.
For $s\ge 1$ the supercurrents  satisfy on-shell the conservation equations
\bea
D^{\a_1} J_{\alpha_1\alpha_2\cdots\alpha_{2s}}=0
~,~~~
\tilde{D}^{\a_1} J_{\alpha_1\alpha_2\cdots\alpha_{2s}}=0
~,~~~
\Longrightarrow
~~~
(\g^\mu)^{\a_1\a_2}\pa_\mu J_{\a_1\a_2\a_3\cdots\alpha_{2s}}=0
~.
\eea
Note that the spin-$1$ supercurrent \cite{Dumitrescu:2011iu,Kuzenko:2011rd}
\begin{equation}
J_{\a\b} =
- 2({\tilde{D}}_{(\a}\tilde{\U}_i) D_{\b)}\U^i
+ 2 \ri\tilde{\U}_i(\g^\mu)_{\a\b} \partial_\mu\U^i
 -2\ri\U^i(\g^\mu)_{\a\b} \partial_\mu \tilde{\U}_i
\end{equation}
contains the stress-energy tensor, the supersymmetry current and the current for the $U(1)_R$ symmetry.
Similarly, the currents $J_{\a_1\cdots\a_{2s}}$ comprise several component HS currents including two series of 
integer and half-integer ones \cite{Nizami:2013tpa}. From here onwards we concentrate on deformations involving 
only the spin zero, $J^{(0)}$, and spin one, $J_{\a\b}$, supercurrents.

The scalar superfield $V\equiv H^{(0)}$ describes an $\cN=2$ vector multiplet
associated with the gauging of the diagonal $U(1)$ symmetry within the flavour $U(N)$ group.
In the context of dS/CFT the scalar components in this multiplet will be dual to the scalar and pseudo-scalar 
currents of the bulk Vasiliev theory. On the other hand the background superfield $H_{\a\b}$ describes 
the gravitational superfield for 3D $\cN=2$ supergravity \cite{Kuzenko:2012qg}. This is the supersymmetric extension 
of a background metric and it contains as components the dual of the spin-two and spin-one currents in the bulk.
The following action 
\bea
S=
\int\rd^3x\rd^4\q\,E\,
\tilde\U_i\re^V\U^i
~,
\label{deformed-U(-N)}
\eea
describes the nonlinear $s=0,1$ deformations of the $U(-N)$ model in a simple manner as a system of 
anti-commuting scalar superfields in an arbitrary off-shell vector multiplet and conformal supergravity background
\cite{Howe:1995zm,Kuzenko:2011xg}. 
Here  $E_M{}^A$ is the supervielbein of the 3D $\cN=2$ background supergeometry
and $E\equiv {\rm Sdet}E_M{}^A$ is its superdeterminant.
The linearization of \eqref{deformed-U(-N)} leads to the Lagrangian in \eqref{Lagrangian-HS} with $s=0,1$.

\subsection{Spin $0\leq s\leq 2$ deformations in components}
\label{deformcomponents}

The above superspace description is potentially abstract for many readers. We therefore give here the action 
in components as well as the details of the supersymmetric backgrounds used in the rest of the paper.

We consider the free chiral multiplets associated to the anti-commuting (anti-)chiral superfields $\U$ and $\tilde\U$.
 In order to study spin 0, 1 and 2 deformations that preserve supersymmetry, 
it suffices to consider the coupling to a general background of 3D new minimal supergravity
\cite{Kuzenko:2011xg} in the presence of a background vector field 
(see also \cite{Closset:2012ru}). 
The explicit ingredients of the action \eqref{deformed-U(-N)} in terms of component fields are presented using the 
notations of \cite{Closset:2012ru,Willett:2016adv}.
 
The field content of off-shell new minimal supergravity in three dimensions comprises as component fields
the metric $g_{\mu\nu}$, the gravitini $\psi_\mu$ and $\tilde{\psi}_\mu$,
the $U(1)_R$ symmetry gauge field $A^{(R)}_\mu$,
a $2$-form gauge field $B_{\mu \nu}$ together with its field strength 
$H=\frac{\ri}{2} \epsilon^{\mu \nu \rho} \partial_\mu B_{\nu \rho}$,
a vector auxiliary field $C_\mu$ and its dual $V^\mu=-\ri\epsilon^{\mu \nu \rho}\partial_\nu C_\rho$.  
The background Abelian vector multiplet comprise a gauge field $A_\mu$ together with its field strength 
$F_{\mu\nu}=\pa_{[\mu}A_{\nu]}$, a scalar field $\s$, the gaugini $\l$ and $\tilde{\l}$, and a second scalar field $D$. 
Note that in Euclidian signature the fields do not need to  be real.

With the background fields that we consider,
the off-shell supersymmetry transformations of the component fields of a chiral multiplet
$\U$ of $R$-charge $q$ are:
\bsubeq\label{rsts} 
\bea
\delta \varphi &=& \sqrt{2} \e \chi
~, 
\\
 \delta \chi&=& \sqrt{2} \e G
- \sqrt{2} \ri \gamma^\mu \tilde{\e} D_\mu \varphi
+ \sqrt{2} \ri \sigma \tilde{\e} \varphi 
+ q \sqrt{2} \ri H \tilde{\e} \varphi
~,
\\
\delta G &=& - \sqrt{2} \ri D_\mu(\tilde{\e} \gamma^\mu \chi) 
-\sqrt{2} \ri \sigma \tilde{\e} \chi
+ 2 \ri  \tilde{\e} \tilde{\lambda} \varphi 
- \sqrt{2} \ri(q-2) H \tilde{\e} \chi
~,
\eea
\esubeq
where the $\e$ and $\tilde{\e}$ are the supersymmetry parameters with 
$U(1)_R$ charge $1$ and $-1$, respectively.
The covariant derivative $D_\mu$ are  gauge and $U(1)_R$ covariant besides including the spin connection.

We consider the case where the $U(1)_R$ charge for the chiral supermultiplet is $q=1/2$. This leads to 
the superconformal theory with Lagrangian 
\bea
{\cal L}
&=& 
D_\mu \tilde{\varphi}  D^\mu \varphi 
+ \tilde{\varphi} \Big(
\frac{1}{8} R 
+\sigma^2  
+ D\Big) \varphi 
+ \ri \tilde{\chi} \gamma^\mu D_\mu \chi 
+ \ri 
\sigma 
\tilde{\chi} \chi 
- \tilde{G} G
~,
\label{chiral-curved}
\eea
where $R$ is the scalar curvature of the background manifold. 
This is the component version of the superspace action \eqref{deformed-U(-N)}.
It is worth mentioning that the quadratic terms in the 
bosonic spinors $\chi$ and $\tilde\chi$ acquire an overall minus sign compared to the case of commuting chiral 
multiplets due to the unusual statistics (compare e.g. with \eqref{free-U(N)}). Including in addition the chiral integral of 
a superpotential $W(\U^I)$ (where we now consider multiple chiral multiplets labelled by the index $I$), its component 
Lagrangian is
\bea
\cL_W=
\frac{\pa W(\varphi)}{\pa\U^J}G^J
+\frac{\pa^2W(\varphi)}{\pa\U^J\pa\U^I}\chi^I\chi^J
~,
\eea
together with its conjugate.
For the interesting case with $\U_\pm$ multiplets mentioned earlier, the classically marginal deformation
$W_{\rm marginal}(\U_\pm)=(\U_+^i\U_-{}_i)^2$ leads to the component action
\bea
\cL_{W_{\rm marginal}}&=&
2(\varphi_+^k\varphi_-{}_k)\big(\varphi_+^iG_{-}{}_i-\varphi_-{}_iG_{+}{}^i\big)
\non\\
&&
+2\big(
(\varphi_+^i\chi_-{}_i+\varphi_-{}_i\chi_+{}^i)^2
-2\varphi_+^i\varphi_-{}_i\chi_+^j\chi_-{}_j
\big)
~.
\label{chiral-curved_interactions}
\eea
Integrating out the auxiliary field $G_{\pm}$ from the sum of \eqref{chiral-curved} and 
\eqref{chiral-curved_interactions} we obtain the classically marginal deformation $V_{\rm marginal}$
\bea
V_{\rm marginal}
&=&
8(\varphi_+^i\varphi_-{}_i)(\tilde\varphi_+{}_j\tilde\varphi_-{}^j)\big(
\varphi_+^k\tilde\varphi_+{}_k
+\varphi_-{}_k\tilde\varphi_-{}^k
\big)
\non\\
&&
+2\big(
(\varphi_+^i\chi_-{}_i+\varphi_-{}_i\chi_+{}^i)^2
-2\varphi_+^i\varphi_-{}_i\chi_+^j\chi_-{}_j
\big)
~,
\label{chiral-curved_interactions_2}
\eea
where all the interaction terms are  double and triple trace operators.

So far we have neglected the gaugini and gravitini
which would couple to spin 1/2 and 3/2 currents in the actions. 
By truncating these modes, 
the invariance of the action require the supersymmetry variation of the background fermions
to be zero. 
By imposing the variation of the gravitini to be zero,
$\delta \psi_\mu=\delta \tilde{\psi}_\mu=0$, one finds the Killing spinor equations:
\bsubeq\bea\label{ckse}
(\nabla_\mu - \ri A^{(R)}_\mu) \e
&=& 
- \frac{1}{2} H \gamma_\mu \e 
- \ri V_\mu \e 
- \frac{1}{2} \epsilon_{\mu \nu \rho} V^\nu \gamma^\rho \e 
~,
\\  
(\nabla_\mu + \ri A^{(R)}_\mu) \tilde{\e} &=& 
- \frac{1}{2} H \gamma_\mu\tilde{\e} 
+ \ri V_\mu \tilde{\e} 
+ \frac{1}{2} \epsilon_{\mu \nu \rho} V^\nu \gamma^\rho \tilde{\e} 
~.
\eea
\esubeq
Requiring the variation of the gaugini to be zero,
 $\d\l=\d\tilde{\l}=0$, one finds the following additional constraints on the background fields
\bsubeq\label{gengt} 
\bea
0&=& \Big(\ri  (D+\sigma H) - \frac{\ri}{2} \epsilon^{\mu \nu \rho} \gamma_\rho  F_{\mu \nu} 
- \ri \gamma^\mu  (\pa_\mu \sigma + \ri V_\mu \sigma) \Big) \e
~,\\
0&=&
  \Big(- \ri  (D +\sigma H) - \frac{\ri}{2} \epsilon^{\mu \nu\rho} \gamma_\rho F_{\mu \nu} 
+ \ri \gamma^\mu (\pa_\mu \sigma -\ri V_\mu \sigma) \Big) \tilde{\e} 
~.
\eea
\esubeq
We refer to \cite{Willett:2016adv} and references therein
for details about the geometrical constraints imposed on a 3-manifold admitting
some residual rigid supersymmetry. The sphere $S^3$ of radius $l$  is a maximally supersymmetric background. 
In this case, besides the metric the only nontrivial field turned on is a constant $H$-flux of the form
$H=\ri/l$.

\subsection{Duality}
\label{dual}

We conjecture that the $\mathcal{N}=2$ supersymmetric extensions of the Euclidean $Sp(N)$ model that we 
constructed above are dual to the $\mathcal{N}=2$ supersymmetric higher-spin gravity theories in de Sitter space 
found in \cite{Sezgin:2012ag}.\footnote{We refer in particular to Section 4.4 of \cite{Sezgin:2012ag} for a detailed 
discussion of these HS theories.}
  
There is a continuous family of Vasiliev theories in de Sitter that depend on the parity-breaking angle $\theta$. In the 
CFT dual, $\theta$ corresponds to a Chern-Simons coupling constant $k$, i.e.
$\theta=\frac{\pi N}{2k}$ \cite{Chang:2012kt}. 
We first consider Vasiliev theories with $\theta=0$. This corresponds to the $k \rightarrow \infty$ limit such that 
$N/k\to0$. 
In this limit we can neglect the Chern-Simons terms since our boundary is simply connected. The dual field theory is 
then free and determined by the symmetry group, i.e.~the amount of supersymmetry and the choice of internal gauge 
symmetry. Its Lagrangian is given in full generality in \eqref{Lagrangian-HS}, and in \eqref{chiral-curved} when 
restricted to low spin deformations.
The symmetry groups of the bulk and boundary theories are the same: 
both theories have $OSp(2|2, 2)$ invariance possessing 8 real supercharges. 
They also have the same higher-spin symmetry,\footnote{The higher spin symmetry of free vector models follows 
from the properties of the Dirac and Laplace operators and hence does not depend on the spin-statistics of the field 
on which these act, see e.g. \cite{Eastwood:2002su} 
and \cite{Howe:2016iqw} for the super-Laplacian case.} which brings the full symmetry generated by spacetime, 
supersymmetry and HS of both theories to $ho(1,1|4,1)$.\footnote{By extending the 
analysis of \cite{Bekaert:2009ud,Bekaert:2010ky} (see also \cite{Tseytlin:2002gz,Segal:2002gd}) to the 
supersymmetric case 
it may be possible to prove directly that the partition function of \eqref{Lagrangian-HS} has $ho(1,1|4,1)$ symmetry.
Along this direction, 
regarding the use of Noether method in superspace see the recent paper \cite{Buchbinder:2017nuc}
for the 4D $\cN=1$ case.}

Sezgin and Sundell \cite{Sezgin:2012ag} note that the counting of dynamical fields of 
$\mathcal{N}=2$ ${\rm dS}_4$ higher spin theory matches that of $\mathcal{N}=2$ ${\rm AdS}_4$ higher spin theory. 
Since the counting of $\mathcal{N}=2$ $U(N)$ currents matches the $\mathcal{N}=2$ ${\rm AdS}_4$ bulk fields, 
mutatis mutandis the counting of $\mathcal{N}=2$ $U(-N)$ currents matches the $\mathcal{N}=2$ ${\rm dS}_4$ bulk 
fields.

The bulk theory contains in particular a scalar and a pseudo-scalar with mass $m^2 = 2/R^2$. This mass allows for 
two different boundary conditions which in the boundary theory correspond to either a free or an interacting CFT 
\cite{Anninos:2011ui,Chang:2012kt}. Here we consider the free boundary theory in which the $U(-N)$ singlet scalar 
operators $\tilde{\varphi}_i \varphi^i$ have dimension $\Delta = 1$ and $\tilde{\chi}_i \chi^i$ have $\Delta = 2$. 
They correspond respectively to the bulk scalar and the pseudo-scalar. This means the former must obey alternative 
boundary conditions for which the modes asymptotically behave as $\sim z$ whereas the latter obeys standard 
boundary (or quantization) conditions, with modes behaving as  $\sim z^2$ asymptotically.\footnote{The dimension 
$\Delta_{\pm}$ of operators in the boundary theory is related to the mass $m^2$ of dual bulk scalars as 
$\Delta_{\pm} = \frac{3}{2} \pm \sqrt{\frac{9}{4} - m^2 R^2}$ where $R$ is the dS radius.}

There are also two towers of integer spin bulk fields or equivalently two towers of boundary integer spin conserved 
currents. One tower is constructed by applying derivatives to $\tilde{\phi}_i \phi^i$ to make traceless spin-$s$ currents. 
The other tower is made analogously from $\tilde{\psi}_i \psi^i$. 

Finally there is one tower of strictly half-integer spin fields built by taking combinations of derivatives of $\psi_{\alpha}^i$ and $\tilde{\phi}_i$ and the conjugate. Fermionic half-integer higher-spin fields in the bulk enter as Grassmann valued arguments in the wave function\footnote{To avoid confusion we note that, unlike in ordinary first quantised mechanics, the wave function in quantum cosmology does not explicitly depend on position, or momenta. Rather, the arguments of wave functions in quantum cosmology are configurations of matter fields and three-geometries which themselves are in general of course functions of position.} \eqref{dSCFT}. The general framework of supersymmetric quantum cosmology was developed long ago, see e.g. \cite{DEath:1986lxx,DEath:1988jyp,DEath:1992wve}. One might be concerned about the physical interpretation of wave functions that include Grassmann valued arguments. However, the wave function is not directly a physical observable. Probabilities for physical observables $\cO$ are computed as
\begin{equation}
<\Psi, \cO \Psi>=\int \cD h \cD A_{s \in \mathbb{N}} \cD A_{s \in \mathbb{N}+1/2} \overline{\Psi}(h,A_{s}) \cO(h,A_{s}) \Psi(h,A_{s}) \Xi(h,A_{s})
~.
\end{equation}
Here we have denoted separately and abstractly the integer and half-integer matter field configurations in the functional integral and we have inserted a density measure $\Xi$. This shows that for bosonic operators $\cO$, since one integrates over the Grassmann valued $ A_{s \in \mathbb{N}+1/2}$ fields, physically meaningful expectation values are ordinary numbers.

We conclude with a remark on interacting theories. We have constructed interacting supersymmetric vector models, with a potential of the form \eqref{chiral-curved_interactions_2}. It is natural to conjecture these play a role in a duality with supersymmetric higher-spin gravity theories in dS with the parity-breaking $\theta$ angle turned on. The analogous AdS/CFT discussion \cite{Chang:2012kt} shows that marginal double and triple trace interactions correspond to generalized `designer gravity' boundary conditions 
\cite{Hertog:2004ns} in the bulk. Interaction terms involving both $\varphi$ and $\chi$ will imply a relation between the 
asymptotic profiles of the bulk scalar and pseudo-scalar. In the context of dS/CFT this means the partition function 
computes the wave function in an unusual basis \cite{Anninos:2012ft}. With $\theta$ turned on, however, also the 
Chern-Simons terms are important \cite{Chang:2012kt,Honda:2017nku}. 
In this regard we expect extended $\cN>2$ supersymmetric models to play an important and interesting 
role, in which the interplay between Chern-Simons and matter dynamics can render the marginal 
deformations exactly marginal. We leave a more precise and complete formulation of dS/CFT for interacting duals to 
future work and turn now to a first exploration of the physics of the duality relating the free theories.

\section{Supersymmetric Minisuperspace}
\label{partition-function}

We now evaluate the partition function of the free $U(-N)$ model \eqref{chiral-curved} for a range of 
supersymmetry preserving bosonic deformations of the theory. We first consider homogeneous mass deformations of the theory. These 
correspond to turning on a supersymmetric vector multiplet background. Next we compute the theory on 
squashed three-sphere boundaries where we will see supersymmetry requires an additional spin-1 deformation be 
turned on. 

Eq. \eqref{dSCFT} shows that dS/CFT relates the sources of the deformations in the partition function to the argument of 
the bulk wave function in the large three-volume limit. The dependence of the partition function on the values of the above sources therefore yields a cosmological measure on a minisuperspace of asymptotically dS configurations. From this measure predictions for semi-local observables in cosmology can be derived through further coarse-graining. The deformations we consider here specify a minisuperspace that comprises a class of anisotropic deformations of de Sitter space (which have squashed sphere future boundaries) and asymptotic dS universes with an early phase of scalar field inflation.

We consider free dual theories only in this section. Their partition functions are inversely related to the partition 
functions of the corresponding $U(N)$ theories with ordinary spin-statistics. By Gaussian integration the latter take the 
form
\begin{equation}\label{inverse}
Z \propto \frac{\det(\text{half-integer spin field eigenvalues})}{\det(\text{integer spin field eigenvalues})}\text{,}
\end{equation}
whereas for the theories with opposite spin statistics we have,
\begin{equation}\label{no-inverse}
Z \propto \frac{\det(\text{integer spin field eigenvalues})}{\det(\text{half-integer spin field eigenvalues})}\text{.}
\end{equation}
This relation is an exact example of the more general close connection between (Euclidean) AdS/CFT and dS/CFT 
evoked in the Introduction. It means that for the purpose of this section we can either work with the original free chiral 
multiplets (see e.g.~\cite{Willett:2016adv}) and use \eqref{inverse} or evaluate directly \eqref{no-inverse}. We illustrate 
both methods below.

Before we proceed with our analysis we should comment on two subtleties. First, in the presence of 
background fields contact terms can appear that render the value of the partition function unphysical \cite{Closset:2012vp}. 
A meaningful dS/CFT duality of the form \eqref{dSCFT} requires at least the norm of the partition function be physical, which is guaranteed if the duals are reflection positive. However we lack a proof of this in the present context\footnote{With more supersymmetry fewer contact terms are possible. For instance the Chern-Simons term $A\wedge d A$ is absent whenever $\mathcal{N}>2$.}. A second subtlety concerns the regularisation of the determinants. On the $U(N)$ side the regularisation is fixed by $\mathcal{N}=2$ localisation \cite{Closset:2012ru}. We assume the same regularisation holds in the $U(-N)$ case. Localisation for the $U(-N)$ theory should parallel the result of the $U(N)$ theory because the off-shell supersymmetry on the three manifolds is the same in the two cases. 

\subsection{Scalar deformations}
\label{sec:scalardeform}

The action of the undeformed supersymmetric $U(-N)$ model \eqref{chiral-curved} on a round $S^3$ boundary is
\begin{equation} \label{chiral-S3}
    S_{chi} = \int d^3x \sqrt{h} \left[ \partial_\mu \tilde{\varphi}_i \partial^\mu \varphi^i 
    + \frac{3}{4 l^2} \tilde{\varphi}_i \varphi^i 
    +\ri \tilde{\chi}_i \slashed{\nabla} \chi^i 
    - \tilde{G}_i G^i\right]\text{.}
\end{equation}
We consider mass deformations of this theory which break conformal symmetry, but preserve 
supersymmetry\footnote{One can also consider supersymmetry breaking mass deformations by leaving $\sigma$ 
and $D$ in \eqref{chiral-curved} independent. Taking $\sigma = 0$ for instance yields the setup of 
\cite{Anninos:2012ft} multiplied by an overall factor coming from the spinors.} by a coupling to 
the vector multiplet $V$. 
First we set the gauge multiplet to a BPS configuration in which only the scalar fields $\s$ and $D$ are nonzero and 
constant. The constraints \eqref{gengt} then imply that a supersymmetric vector multiplet background has $D=-\ri \s/l$. 
Substituting this condition in \eqref{chiral-S3} yields the following mass deformation,
\begin{equation}
\label{imaginarymassdeform}
    \mathcal{L}_{mass} = \left(\s^2 - \ri \frac{\s}{l} \right) \tilde{\varphi}_i \varphi^i +\ri \s \tilde{\chi}_i \chi^i
    \text{ .}
\end{equation}
where $l$ is the radius of the sphere.

We see that the deformation gives masses to both the scalars and the spinors in the boundary theory. 
As discussed above the component scalar current $j_+^{(0)}=\tilde{\varphi}_i \varphi^i$ is dual to a bulk scalar
whereas the current $j_-^{(0)}=\tilde{\chi}_i \chi^i$ constructed out of the spinor fields is dual to the bulk pseudoscalar. 
Supersymmetry implies both fields are coupled.

The independent complex scalar field $\s$ plays the role of an external, constant mass parameter in the deformed 
theory \cite{Willett:2016adv}. It therefore enters as a source in the partition function. In the bulk $\s$ corresponds to 
the coefficient of the subleading term $\sim z^2$ in the asymptotic profile of the scalar. The phase of $\s$ is determined from the requirement that 
the theory must have a well-defined, asymptotically classical, 
real de Sitter structure \cite{Hartle2012b}\footnote{A different but equivalent way to see this is that any other `reality' 
condition on $\sigma$ would imply the bulk scalar to behave as a ghost, and the resulting wave function to be 
ill-defined.}. 
For $m^2=2/R^2$ scalars this implies in particular we must take $\s$ imaginary \cite{Anninos2014,Hertog:2016lik}. 
Hence the asymptotic profile of the bulk scalar has a real leading term and an imaginary subleading term, precisely 
what one expects for scalar fields in the 
Euclidean Hartle-Hawking vacuum. For the susy deformations we consider, the boundary conditions on the pseudo-scalar are completely fixed by those on the scalar. We are thus led to consider the following mass deformation $\sigma=\ri m$, 
with $m$ real,
\begin{equation}
\label{massdeform}
    \mathcal{L}_{mass} = \left[-m^2 + \frac{m}{l} \right] \tilde{\varphi}_i \varphi^i 
    -m \tilde{\chi}_i \chi^i\text{ .}
\end{equation}

To evaluate the partition function as a function of $m$ we must first find the eigenvalues of the Laplace and Dirac 
operators on $S^3$. These are well-known and can for instance be found in 
\cite{Willett:2016adv} (see \cite{Samsonov:2014pya} for a derivation in superspace). 
Setting $l=1$ the scalar eigenvalues are given by
\begin{equation}
\lambda_n = n(n+2) + 3/4 - m^2 + m\text{,}
\end{equation}
with $n=0,1,2,..$ and degeneracy $(n+1)^2$. Note that the first eigenvalue is negative when $m<-1/2$. 
The eigenvalues for the spinor are
\begin{equation}
\lambda_n^\pm = \pm \Big(n+\frac{1}{2}\Big) - m\text{,}
\end{equation}
with $n=1,2,3,..$ and degeneracy $n(n+1)$. There are negative eigenvalues for all values of $m$. 
However, each eigenvalue has an even degeneracy. Since eigenvalues are raised to the power of their degeneracy 
in the partition function this means the spinor contribution need not necessarily lead to divergences.

\begin{figure}
\centering
\includegraphics[scale=0.55]{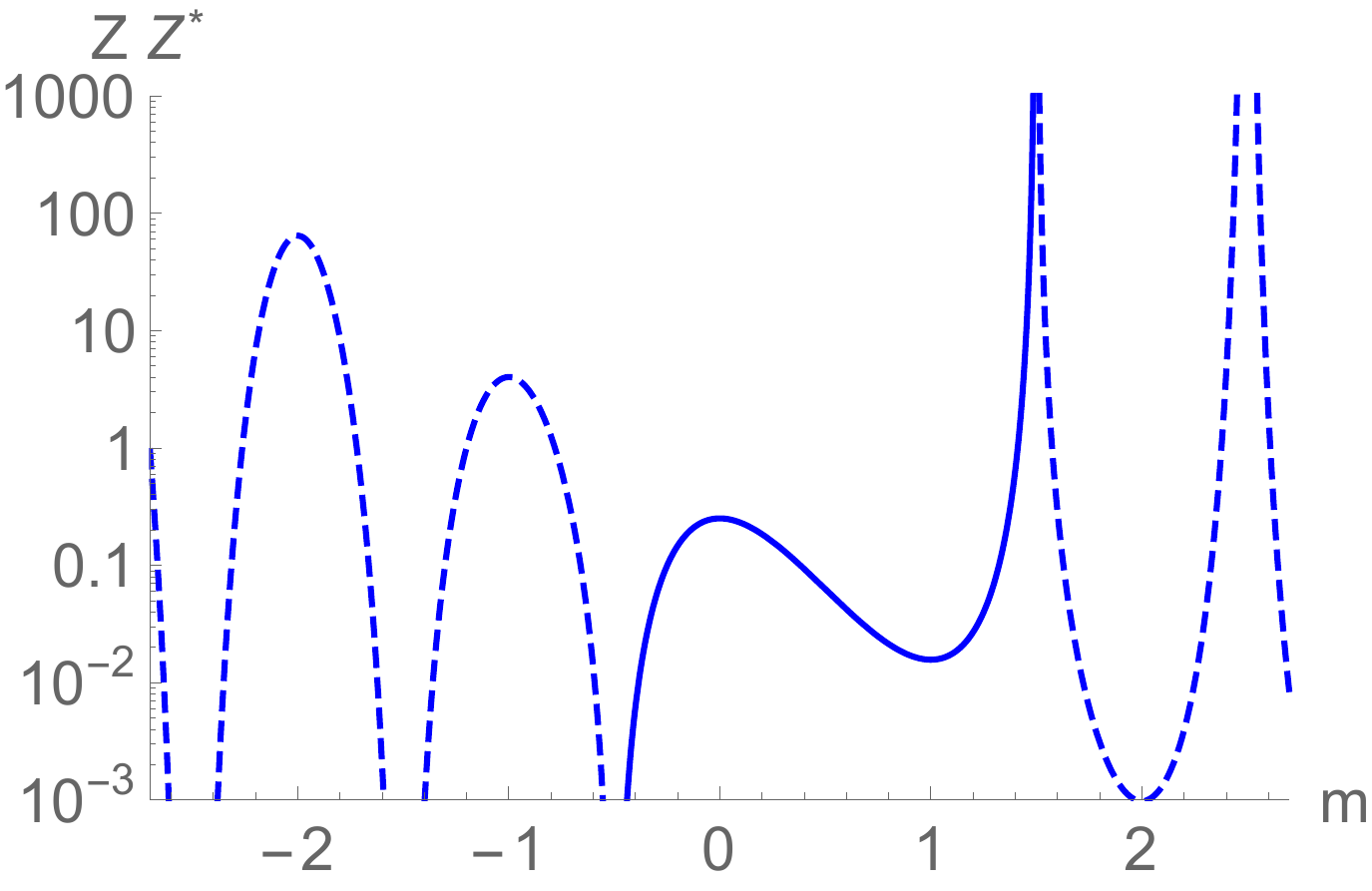}
\includegraphics[scale=0.55]{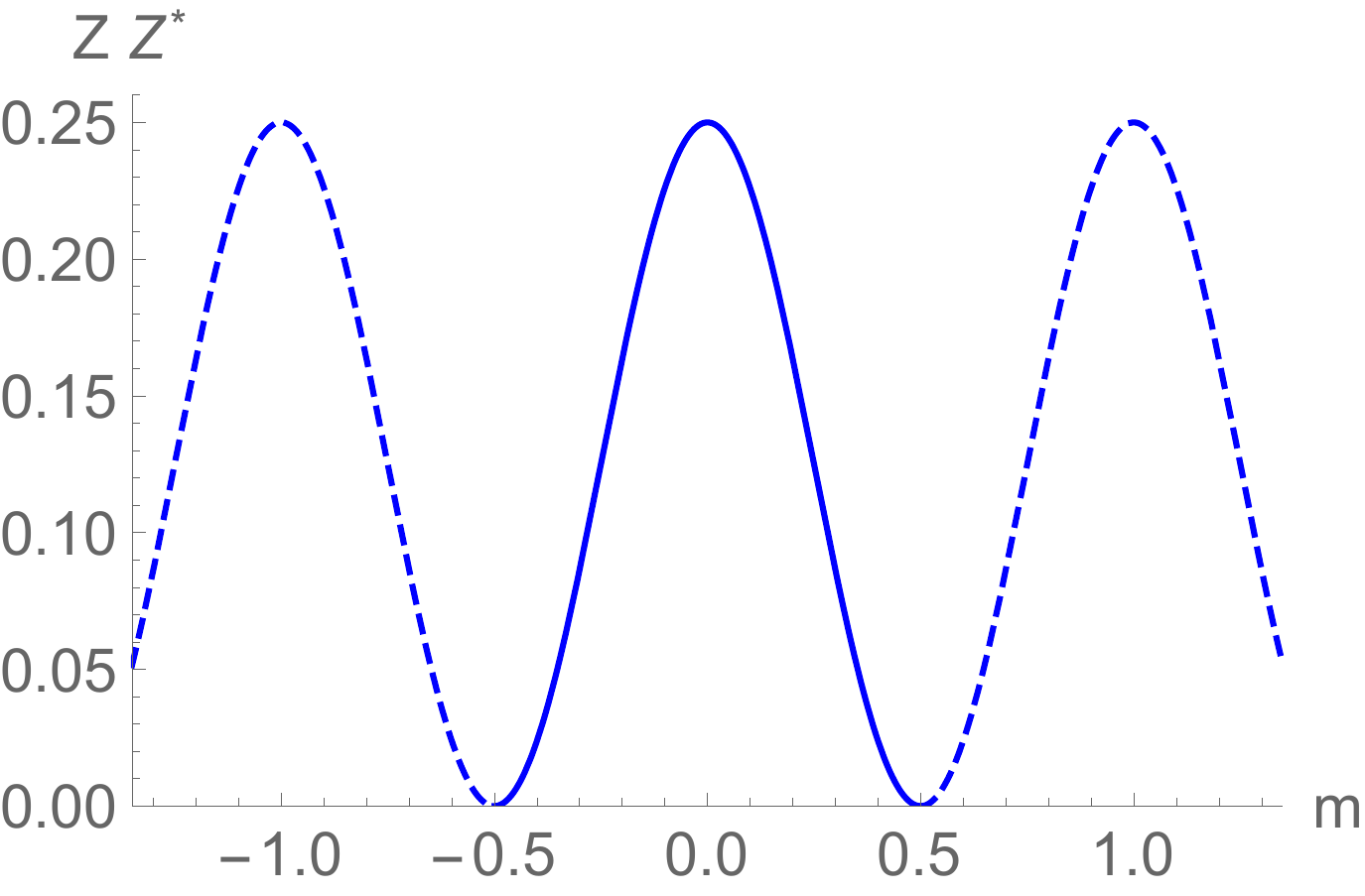}
\caption{The response of $Z Z^*$ to various mass deformations for two chiral multiplets with reversed spin-statistics. 
We conjecture this corresponds to the Hartle-Hawking wave function in a homogeneous isotropic minisuperspace 
model in HS gravity. \textbf{Left:} Both multiplets are in the same representation of the gauge multiplets inducing the 
mass deformation \cref{massdeform} with parameter $m$. \textbf{Right:} the two multiplets are in opposite 
representations; one has mass parameter $m$ and the other $-m$ under \cref{massdeform}. }
\label{fig:cosy}
\label{fig:asymmy}
\end{figure}

With the eigenvalues at our disposal it is straightforward to evaluate the Gaussian integrals defining the partition 
function $Z[m]$.
See \cite{Willett:2016adv} and references therein for a pedagogical description of the case with standard statistic.
 For a single multiplet in the theory with reversed spin-statistics we get 
\begin{equation}
\label{Zmassderform}
Z[m] = \prod_{j=0}^{\infty} \left[\frac{j + m + 1/2}{j-m+3/2}\right]^{j+1} \text{.}
\end{equation}
To numerically perform the product in \eqref{Zmassderform} we found it  convenient to first calculate the second 
derivative of the free energy. Fig \ref{fig:asymmy}(a) shows the resulting distribution $Z Z^*$ as a function of the mass 
parameter $m$. 
One sees it has a local maximum at zero deformation, corresponding to the amplitude of pure dS space in higher-spin 
gravity. This is in accordance with general field theory results such as the F--theorem.\footnote{Had we instead taken 
$\sigma$ real we would have found a pathological wave function even for small deformations. In particular the 
distribution would have had a $\cosh(m)$ behavior. Hence pure dS would have been a local minimum of the 
distribution, leading to two-point functions characteristic of an unstable ghostlike theory.} 
For negative masses the distribution closely resembles that found in \cite{Anninos:2012ft}. In particular it goes to zero 
as $m \rightarrow -1/2$. For more negative masses the holographic wave function is zero. This is because the path 
integral defining the partition function diverges in deformed (free) theories for which the scalar operator in the action 
has a negative eigenvalue. Hence the form \eqref{Zmassderform} no longer holds. Therefore we show the behavior 
following from \eqref{Zmassderform} with dotted curves in Fig \ref{fig:asymmy}(a). 

For positive masses the distribution diverges as $m \rightarrow 3/2$. Whether or not this renders the wave function 
non-normalizable and ill-defined may depend on the integration measure. However this behavior can be improved by 
adding a flavor symmetry. This can be done by placing $N$ chiral multiplets in the fundamental and $N$ in the 
anti-fundamental of the gauge multiplet,
as for the case with $\U_{+}^i$ and $\U_{-}{}_i$ described in section \ref{sec:deform}. 
In the BPS configuration of the gauge multiplet, this means half of the fields have opposite mass $m$. 
Fig \ref{fig:cosy}(b) shows the resulting distribution $ZZ^*[m]$ which is of the form of a cosine squared. 
The wave function now has support on the interval $ \vert m \vert  \leq 1/2$ only. For larger values it is zero. 
The distribution is well-behaved over the entire configuration space and has a global maximum at $m=0$. 

The addition of a flavor symmetry may appear ad hoc, but it is not. In \cite{Chang:2012kt}, it is precisely the addition 
of such flavor symmetry that relates the model to string theory. Imposing this constraint from string theory in our setup 
appears to enhance the stability of de Sitter space. 

The distribution $ZZ^*[m]$ specifies the no-boundary wave function $\Psi[A_+^{(0)},A_-^{(0)}]$ in a supersymmetric 
minisuperspace model consisting of homogeneous, asymptotically de Sitter universes in HS gravity with a 
scalar $A_+^{(0)}$ 
and a pseudo-scalar $A_-^{(0)}$ turned on. The histories therefore exhibit an interior region with 
scalar field driven inflation. The upper bound on the deformation $m$ might mean the bulk potential is such that there 
is a maximum number of efolds of scalar field inflation.

\subsection{Squashings and Vector Fields}
\label{sec:squashings}

Next we consider deformations of the background metric which preserve four real supercharges. In particular we 
consider the partition function of the free $U(-N)$ model \eqref{chiral-curved} on homogeneous squashings of the three-sphere that are 
characterized by a squashing parameter $v>0$. The metric can be written as
\begin{equation}
ds^2 = d\theta^2 + \sin^2 \theta d \phi^2 + \frac{1}{v^2}(d\psi + \cos \theta d\phi)^2 \ 
\end{equation}
where $\theta$, $\phi$, $\psi$ are the Euler angles on $S^3$ such that $\theta \in [0,\pi]$, $\phi \in [0,2\pi]$ and 
$\psi \in [0,4\pi]$. Supersymmetry requires that we turn on a background $U(1)_R$ symmetry gauge field on 
squashed backgrounds. 
In the $U(N)$ model with standard spin-statistics, this is given by the one-form 
\cite{Imamura:2011wg,Martelli:2013aqa}
\begin{equation}\label{gauge}
A^{(R)} = - \frac{1}{2 v^2}\sqrt{1 - v^2}(d\psi + \cos \theta d\phi)\text{.}
\end{equation}

In this section we evaluate the wave function with asymptotic dS boundary conditions using the inverse of the partition 
function of deformations of the $U(N)$ theory. As before the phases of the sources are specified by the condition that 
the wave function is defined on a real configuration space that is asymptotically dS \cite{Hertog2011}. The gauge field $A^{(R)}$ 
is the boundary value of a bulk gauge field $\vec A$ in the dual AdS theory. To evaluate the wave function in the dS 
domain $\vec A$ must be asymptotically imaginary, because the original AdS scale factor $a$ is also imaginary in the 
dS domain. Taken together this yields asymptotically real frame fields $\vec A/a$ on the dS side \cite{Hartle:2013vta}. 
Hence we ought to compute the partition function for purely imaginary values of the source $A^{(R)}$. This selects the 
range $v \geq 1$ in \eqref{gauge}. Note that the round three-sphere corresponds to $v=1$. On this background, the 
gauge field is appropriately turned off. 

The free energy of AdS dual $U(N)$ theories on squashed boundaries was computed in 
\cite{Imamura:2011wg,Martelli:2013aqa,Willett:2016adv,Nishioka:2013gza}. 
From the relations \eqref{inverse} and \eqref{no-inverse} it follows that it suffices to change its overall sign to find the 
free energy of the dS dual theories with reversed spin-statistics. In the $ v \geq 1$ domain this yields the following 
partition function (for a single multiplet),
\begin{equation} \label{Zgauge}
Z(v) = 
\text{Exp}\left[
-\int_{0}^{\infty} \frac{dx}{2x} \left( \frac{\sinh(x/v)}{\sinh\left(x \left(v^{-1}\pm \sqrt{v^{-2}-1}\right)\right)
\sinh\left(x/\left(v^{-1}\pm \sqrt{v^{-2}-1}\right)\right)} - \frac{1}{v x}\right)\right]\;.
\end{equation}

\begin{figure}
    \centering
    \includegraphics[scale=0.6]{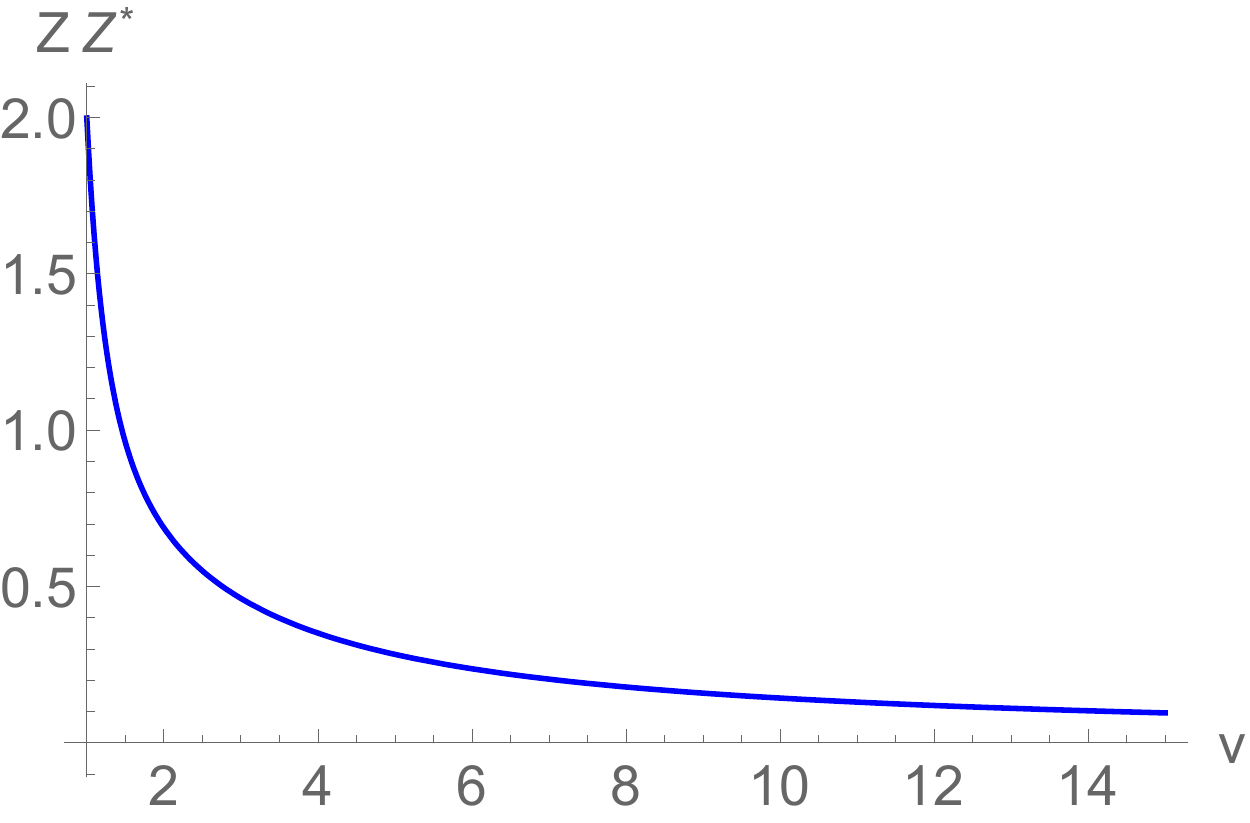}
    \caption{The holographic Hartle-Hawking wave function in higher-spin gravity in the 
one-dimensional minisuperspace of anisotropic but homogeneous deformations of de Sitter coupled to a gauge field, parameterized by $v$, for a single multiplet.}
    \label{fig:foursusysquashing}
\end{figure}

This specifies the large three-volume limit of the no-boundary wave function in higher-spin gravity in the 
one-dimensional minisuperspace of anisotropic deformations of dS coupled to a gauge field. 
Fig. \ref{fig:foursusysquashing} shows the resulting distribution is well-behaved and normalizable with a global 
maximum at pure dS space.
In particular, the constraints implied by supersymmetry -- together with a careful analysis of the asymptotic 
structure -- appear to eliminate the usual problem that gauge fields in supergravity theories on de Sitter backgrounds 
are ghosts.

To conclude we comment on deformations that are non-supersymmetric squashings of chiral multiplets where one 
does not turn on a background gauge field. The free energy of such deformed theories was computed in 
\cite{Bobev:2017asb} in the context of AdS duals. For free chiral multiplets this can be interpreted in the context of 
dS/CFT by inverting the partition function as we discussed. This yields a well-behaved distribution in which 
undeformed dS space seems stable, at least against small deformations. This indicates that for $v \geq 1$ the 
contributions to $Z$ due to the chiral multiplet and the background gauge field are separately stable under 
squashings. On the other hand, for $v<1$ there is a clear difference. The translation to dS of the results of 
\cite{Bobev:2017asb} shows that $v=1$ is a local maximum for non-supersymmetric squashings even when one 
includes the $v<1$ regime. By contrast this would not be the case if one were to extend our result for supersymmetric 
squashings to $v <1$, which would yield a distribution that diverges as $v \rightarrow 0$. This is precisely what one 
expects because under the boundary conditions corresponding to $v<1$, which we did not take, the vector field is 
ghost-like \cite{Hartle:2013vta}. It is interesting that the $v<1$ regime covers the entire region of configuration space 
where the Yamabe invariant of the conformal boundary is negative. The fact that the holographic wave function has no 
support in this region once gauge fields are taken in account lends further support to the conjecture 
\cite{Hawking:2017wrd} that the holographic measure strongly suppresses conformal boundaries far from the 
round conformal structure.

\section{Speculations on a triality with exotic string theories}
\label{sec:string}

In the context of ordinary AdS/CFT it has been argued \cite{Chang:2012kt} 
that a triality relates type IIA string theory on ${\rm AdS}_4\times {\rm CP}_3$ (possibly with some $B$-field flux) 
to specific Vasiliev theories in ${\rm AdS}_4$ and Chern-Simons vector models. These
type IIA backgrounds  lift to 11-dimensional supergravity on 
${\rm AdS}_4\times {\rm S}^7/{\rm Z}_k$ (possibly with torsional flux added). 

Our results suggest this triality can be `Wick rotated' and thus generalized to a dS context where the third party would 
be the exotic M-theory ${\rm M}^-_{9,2}$ in a ${\rm dS}_{4}\times {\rm AdS}_7/{\rm Z}_k$ background constructed by 
Hull \cite{Hull:1998vg}.\footnote{See \cite{Dijkgraaf:2016lym} for this notation, where the possible existence of exotic 
string theories was also connected to supergroup-gaugings and negative branes. Negative branes naturally lead to 
signature changes between different regions of spacetime and thereby connect AdS and dS. Hence they may play a 
role in a microscopic description of the no-boundary wave function.} 
This exotic M-theory has two time directions and the M2 brane has a Euclidean world volume. 
In flat space its near-horizon geometry is ${\rm d S}_{4}\times {\rm AdS}_7$. 
The Vasiliev theories in dS would then be the tensionless limit of this theory.

A triality of this kind would resonate with the intuition recalled in the Introduction that exotic string theories in dS can 
be well-behaved despite having ghosts in the supergravity limit. Hull has argued that the massive string states 
render the starred string theories ghost-free \cite{Hull:1998vg}. The Vasiliev limit which features in the triality precisely 
corresponds to the limit where the massive string tower becomes massless. It is therefore plausible that Hull's 
`ghost-exorcism' shows up at the classical higher-spin level. The fact that the minisuperspace wave function computed in Section \ref{partition-function} is well behaved provides some holographic support for this.   

We conclude with an illustration of how this triality might be realized more explicitly. For this purpose let us regard the 
above setup from a type IIB viewpoint. It is well known that ordinary ABJ(M)-theory emerges in the IR from specific 
brane configurations (cf. Table \ref{table:ABJM}(a)). In Table \ref{table:ABJM}(b) we configure branes in an analogous 
manner in the exotic ${\rm IIB}^{-+}_{(9,1)}$ theory. The $D3$-branes have become Euclidean while the $D5$-branes 
are Lorentzian. If one can argue that strings between the Euclidean $D3$ and the ordinary $D5$ branes have 
reversed spin-statistics, then the resulting 3D field theories have limits in which they turn into the Euclidean $U(-N)$ 
field theories we have constructed. For example, separating the Euclidean $D3$ and ordinary $D5$ branes would 
then correspond exactly to the mass deformation in Section \ref{sec:scalardeform} with an additional flavor symmetry that 
gives half the fields opposite masses.\footnote{As an aside we note that it was argued \cite{Parikh:2002py} that the 
near-horizon geometry of the Euclidean branes in exotic string theories does not involve de Sitter space but rather 
elliptic de Sitter space (where the antipodal points are identified). This in turn appears to resonate with 
\cite{Hartle:2011rb} where it was shown that the physical arrow of time in asymptotic dS histories predicted by the 
Hartle-Hawking wave function reverses near the dS throat.}

\begin{table}[]
\begin{tabular}{l|llllllllll}
    & 0 & 1 & 2 & 3 & 4 & 5 & 6 & 7 & 8 & 9 \\ \hline
$NS5$ &  $\bullet$  & $\bullet$  & $\bullet$  & $\bullet$  & $\bullet$  & $\bullet$  &    &   &   &   \\
$D3$  & $\bullet$  & $\bullet$  & $\bullet$  &   &   &   & $\bullet$  &   &   &   \\
$D5$  & $\bullet$  & $\bullet$  & $\bullet$  & $\bullet$  & $\bullet$  &   &   &   &   & $\bullet$ 
\end{tabular} 
\qquad
$\Longrightarrow$
\qquad
\begin{tabular}{l|llllllllll}
    & 0 & 1 & 2 & 3 & 4 & 5 & 6 & 7 & 8 & 9 \\ \hline
$ENS5$ &   & $\bullet$  & $\bullet$  & $\bullet$  &  $\bullet$ &  $\bullet$  &  &   &   & $\bullet$  \\
 $ED3$  &   & $\bullet$  & $\bullet$  & $\bullet$  &   &   & $\bullet$  &   &   &   \\
$D5$  & $\bullet$  & $\bullet$  & $\bullet$  & $\bullet$  & $\bullet$  & $\bullet$  &   &   &   &  
\end{tabular}
\caption{\textbf{Left table}: ABJ(M) setup in IIB string theory. The Lorentzian 3D CFT is in the $012$ direction. 
\textbf{Right table}: Analogous setup to ABJ(M) in ${\rm IIB}^{-+}_{(9,1)}$ with Euclidean $D$3 branes. 
The Euclidean 3D CFT is in the $123$ direction.}  \label{table:ABJM}
\end{table}

\section{Discussion}
\label{sec:discussion}

We have constructed supersymmetric Euclidean vector models in three dimensions with reversed spin-statistics. We conjecture these are holographically dual to specific supersymmetric Vasiliev theories in four-dimensional 
de Sitter space. 

We have begun to explore this duality by computing the partition function for a range of scalar, vector and tensor deformations that preserve supersymmetry. The duality asserts this specifies the Hartle-Hawking wave function in a supersymmetric minisuperspace consisting of anisotropic bosonic deformations of dS with scalar and vector matter. We found the wave function is globally peaked at de Sitter, with a low amplitude for strong deformations. This suggests that 
supersymmetric de Sitter space is stable in higher-spin gravity and in particular free from the usual ghosts. 

An important generalization of our analysis concerns the calculation of the CFT partition function for deformations sourcing half-integer spin fields. Free spin-$1/2$ and spin-$3/2$ fields in de Sitter space have been studied in \cite{Anguelova:2003kf}, which also derives a number of general properties of dual CFTs. In particular, \cite{Anguelova:2003kf} identifies the general form of the boundary-boundary two-point function of a bulk spinor, for an arbitrary spacetime dimension and mass, and up to a constant factor. At the time, however, no CFT dual was known to compare this with. Our model provides a concrete setup in which this can be done. The supersymmetric HS bulk theory we consider includes a specific nonlinear, interacting extension of the free spinors considered in \cite{Anguelova:2003kf}. Specializing the general result for the two-point function in \cite{Anguelova:2003kf} to the massless case in four bulk dimensions yields the following structure for the 3D boundary-boundary two-point function,
\begin{equation}
\label{spinor2pt}
<\mathcal{O}_{1/2}(x)\tilde{\mathcal{O}}_{1/2}(x')> = \text{const} \,\frac{\gamma \cdot (x-x')}{(x-x')^4}  
~.
\end{equation}
This can be compared with our CFT result. On the CFT side we have, $\mathcal{O}_{1/2} \propto \tilde{\varphi} \chi$ and analogously for $\tilde{\mathcal{O}}_{1/2}$.
The form of the $\mathcal{O}_{1/2}$ two-point function is determined up to a constant factor by conformal symmetry. 
Since $\mathcal{O}_{1/2}$ has conformal dimension $\Delta=3/2$, one exactly obtains \eqref{spinor2pt}.\footnote{The analogous bulk and boundary computations in the AdS/CFT case are given in \cite{Henningson:1998cd}.}
Therefore our CFT result matches the general structure of the fermionic two-point functions derived from a bulk analysis in \cite{Anguelova:2003kf}. This provided evidence that our proposed dS/CFT duality holds in the half-integer spin sector. 

Deformations of the boundary theory that correspond to turning on spinors in the bulk are couplings of  
$\mathcal{O}_{1/2}$ and $\tilde{\mathcal{O}}_{1/2}$ to the background gaugini $\lambda$ and $\tilde{\lambda}$ 
as described in \cref{deformcomponents}, although we chose to set these terms to zero. Note that the reality 
properties we have imposed on the background scalars $\sigma$ and $D$ imply specific reality properties on the 
background gaugini $\lambda$ and $\tilde{\lambda}$ through the supersymmetry variation equations.
Supersymmetry thus relates the two-point function \eqref{spinor2pt} to other two-point functions containing integer spin currents. We leave a more detailed analysis of the complete partition function in supersymmetric minisuperspace beyond the level of the two-point function, which one would expect also determines the sign of the latter, to future work.

Our results open up new ways to develop dS/CFT further. First, it would be very interesting to demonstrate the 
absence of ghosts directly in Vasiliev gravity in de Sitter. We have mainly focused on the duality for free theories, 
but it would also be interesting to formulate the duality for supersymmetric interacting theories and to clarify in particular 
how the Chern-Simons terms  and extended $\cN>2$ supersymmetry enter. Supersymmetry will also enable one to introduce in dS/CFT new and powerful calculational techniques, such as localisation, that have led to important advances in the context of AdS/CFT in recent years.

Finally, it would be especially interesting to understand whether the supersymmetric higher-spin theories featuring in 
our duality are indeed the tensionless limit of Hull's exotic string theories
in their de Sitter vacua. This would establish a triality similar to the one put forward in \cite{Chang:2012kt}, which in 
turn may lend support to the conjecture \cite{Hull:1998vg} that the massive string states render
 the starred string theories ghost-free. This would constitute a first step toward the generalization of the 
 HS realisations of dS/CFT to other, more realistic theories in dS.

\section*{Acknowledgements}
\noindent
We thank Dionysios Anninos,  Nikolay Bobev, Frederik Denef, Chris Hull, Ruben Monten, Yi Pang
Antoine Van Proeyen and Yannick Vreys for helpful 
discussions. We furthermore thank Nikolay Bobev
and Yi Pang for crucial remarks on a first draft. GV thanks Columbia University 
for its hospitality during part of this work. This work is supported in part 
by the European Research Council grant no. 
ERC-2013-CoG 616732 HoloQosmos, 
the FWO Odysseus grant G.0.E52.14N,
the Interuniversity Attraction Poles Programme initiated by the Belgian Science Policy (P7/37)
 and the C16/16/005 grant of the KULeuven.

\bibliographystyle{utphys}
\bibliography{dSCFT}
\end{document}